\documentclass{article}
\usepackage{amsmath}
\usepackage{bm}
\usepackage[utf8]{inputenc}
\usepackage{epsfig}
\usepackage{float}

\title{Music Genre Classification with Paralleling Recurrent Convolutional Neural Network }
\author{Lin Feng, Shenlan Liu, Jianing Yao}
\date{December 2017}

\begin{document}

\maketitle
\begin{abstract}
    Deep learning has been demonstrated its effectiveness and efficiency in music genre classification. However, the existing achievements still have several shortcomings which impair the performance of this classification task. In this paper, we propose a hybrid architecture which consists of the paralleling CNN and Bi-RNN blocks. They focus on spatial features and temporal frame orders extraction respectively. Then the two outputs are fused into one powerful representation of musical signals and fed into softmax function for classification. The paralleling network guarantees the extracting features robust enough to represent music. Moreover, the experiments prove our proposed architecture improve the music genre classification performance and the additional Bi-RNN block is a supplement for CNNs.
\end{abstract}

\section{Introduction}
With the extensive utilization of various music platforms, an increasing number of music is widely spread, which causes chaos for audiences and those platforms to organize these music. Furthermore, it's impossible to organize and distinguish such a large number of music by manual efforts. Therefore, how to construct a convenient way to deal with this problem is of vital importance but challenging. Most of state-of-the-art methods aim to classify the music genre which is a top-level label on music to help audiences to categorize and describe various music . \cite{tzanetakis2002musical} Meanwhile, exact classification on music genre is crucial for music platforms to organize music into different groups. For this reason, classification on music genre has attracted widely attentions in the field of music information retrieval (MIR) \cite{shawe2005investigation}\cite{west2005finding}.

As two crucial components for music genre classification, feature extraction and classifier learning may greatly influence the performance of most classification systems \cite{Duda2000Pattern}.Feature extraction concentrates on exploring suitable representations of samples which are expected to be classified in terms of feature vectors or pairwise similarities \cite{fu2011survey}. After feature extraction, features and representations of music are fed into a classifier, which aims to map feature vectors into different music genres. Baniya et al. \cite{baniya2014novel} adopt timbral texture features (i.e. Mel-frequency Cepstral Coefficient) and rhythm content features like beat histogram (BH) \cite{tzanetakis2002musical} to represent music signals. Then, they combine Extreme Learning Machine (ELM) \cite{huang2006extreme} with bagging \cite{breiman1996bagging} as a classifier. Arabi et al. \cite{arabi2009enhanced} draw chord features and chord progression information into feature extraction. In addition, by utilizing Support Vector Machine (SVM), they proved chord features in conjunction with low-level features \cite{fu2011survey} can provide higher classification accuracy. The state-of-the-art achievement is reported by Sarkar et al. \cite{sarkar2015music}, which employs Empirical Mode Decomposition (EMD) for signal component extraction and depends only on pitch based features. Even though all methods above achieve good performance in some certain situations, these hand-craft features cannot avoid some fatal disadvantages. The hand-craft features extraction from music signals need some complex process, thus it requires researchers to possess expertise in the musical domain. Furthermore, features which extracted for one certain task lack universality since they may have poor performances in other tasks. 

In recent years, deep learning, especially Convolutional Neural Networks (CNNs) have been utilized in various image classifications successfully. \cite{wei2016hcp}\cite{ciresan2011flexible} Meanwhile, Sander et al. \cite{dieleman2014end} prove that comparing with normal images, spectrograms of music audio can also achieve good performance with CNNs. Under this circumstance, there is a growing tendency of learning robust feature representations from spectrograms of music with CNNs \cite{li2010automatic}\cite{zhang2016improved}. In contrast with traditional methods, CNNs provides an end-to-end training architecture which combine feature extraction with music classification in one stage. And multiple works based on CNNs have shown their superiorities for music genre classification. 

But it is worth noticing, different from ordinary images, spectrograms of music have heavily sequential relationships inside. However, the existing music genre classifications with CNNs are not able to model the long-term temporal information in spectrograms of music data. As we all know, Recurrent Neural Networks \cite{elman1990finding} (RNNs) can model long-term dependencies like music structure or recurrent harmonies \cite{pons2016experimenting} which are significant for music classification. To address all limitation mentioned above, we propose a hybrid learning architecture named Paralleling Recurrent and Convolutional Neural Network (PRCNN), which consists of a CNN block and a Bidirectional Recurrent Neural Network (Bi-RNN) block \cite{schuster1997bidirectional}. The main contribution of our proposed architecture is that the hybrid structure models not only spatial features but also temporal frame orders of music data, which are greatly complementary to music genre classifications comparing with simple CNNs.

The rest of this paper are organized as follows. In Section 2, we retrospect related work of music genre classification and carefully analyze their contribution as well as limitation. Section 3 describes the construction of our proposed hybrid architecture PRCNN for music genre classification in detail. In Section 4, we implement various experiments based on several datasets and demonstrate the validity of our proposed architecture PRCNN. Finally, we draw a conclusion and present some future work in Section 5.

\section{Related work}
Music genre classification is a widely studied area in Music Information Retrieval for categorizing and describing enormous amount of music \cite{tzanetakis2002musical}. Various researches indicate extracting representative features from music signals can heavily improve the performance of classifications. Thus, most existing works focus on extracting robust features to represent music in order to improve the music genre classification performances. Motivated by the success of computer vision \cite{lawrence1997face}, CNNs have also attracted much attention in the field of music genre classification. By training an end-to-end architecture, CNNs have powerful capacities to represent various music with higher-level features. In addition, CNNs require less engineering effort and prior knowledge of one certain field. Li et al. \cite{li2010automatic} declare the variations of musical patterns with a certain transformation such as, Fast Fourier Transform (FFT) and Mel-frequency Cepstral Coefficient (MFCC), are similar to images which work well with CNNs in image classifications \cite{ciresan2011flexible}. Moreover, they prove CNNs are feasible alternates to extract musical patterns features automatically. Although their work brings opportunities to displace hand-craft features, the experimental results, however, show the proposed structure is not robust enough to make testing data perform as excellently as training data. Zhang et al. \cite{zhang2016improved} proposed two networks to improve the performance of music genre classification with CNNs. In order to offer more statistical information to the following layers, max- and average-pooling are operated in conjunction across the entire time axis in one of networks. Tending to improve the accuracy from increased depth, they utilize shortcut connections inspired by residual learning \cite{he2016deep} in another network. The performances of two CNNs are both demonstrated to be improved contrast with previous results based on GTZAN \cite{tzanetakis2002musical} dataset. However, as mentioned in previous section, musical patterns have some temporal relationships which are crucial for music genre classifications but will be dropped in CNNs. For this reason, Choi et al. \cite{choi2016convolutional} design a hybrid model named convolutional recurrent neural network (CRNN), which CNNs and RNNs are exploited as features extractor and temporal summarizer, respectively. Comparing with three existing CNNs, CRNN is demonstrated to improve the performance of music classification via learning more temporal information. But this hybrid model also have its limitation which impair the performance of music classification. Even though CRNN has RNNs to be the temporal summarizer, it can only summarize temporal information from the output of CNNs. Obviously, the temporal relationships of original musical signals are not preserved during operations with CNNs. 

To preserve both spatial features and temporal frame orders of original music signals, we carefully design the hybrid model which consists of paralleling CNN and Bi-RNN blocks. In next section, we will describe our proposed hybrid architecture for music genre classification in detail.

\section{methodology}
\begin{figure}
    \begin{center}
        \includegraphics[width=10cm]{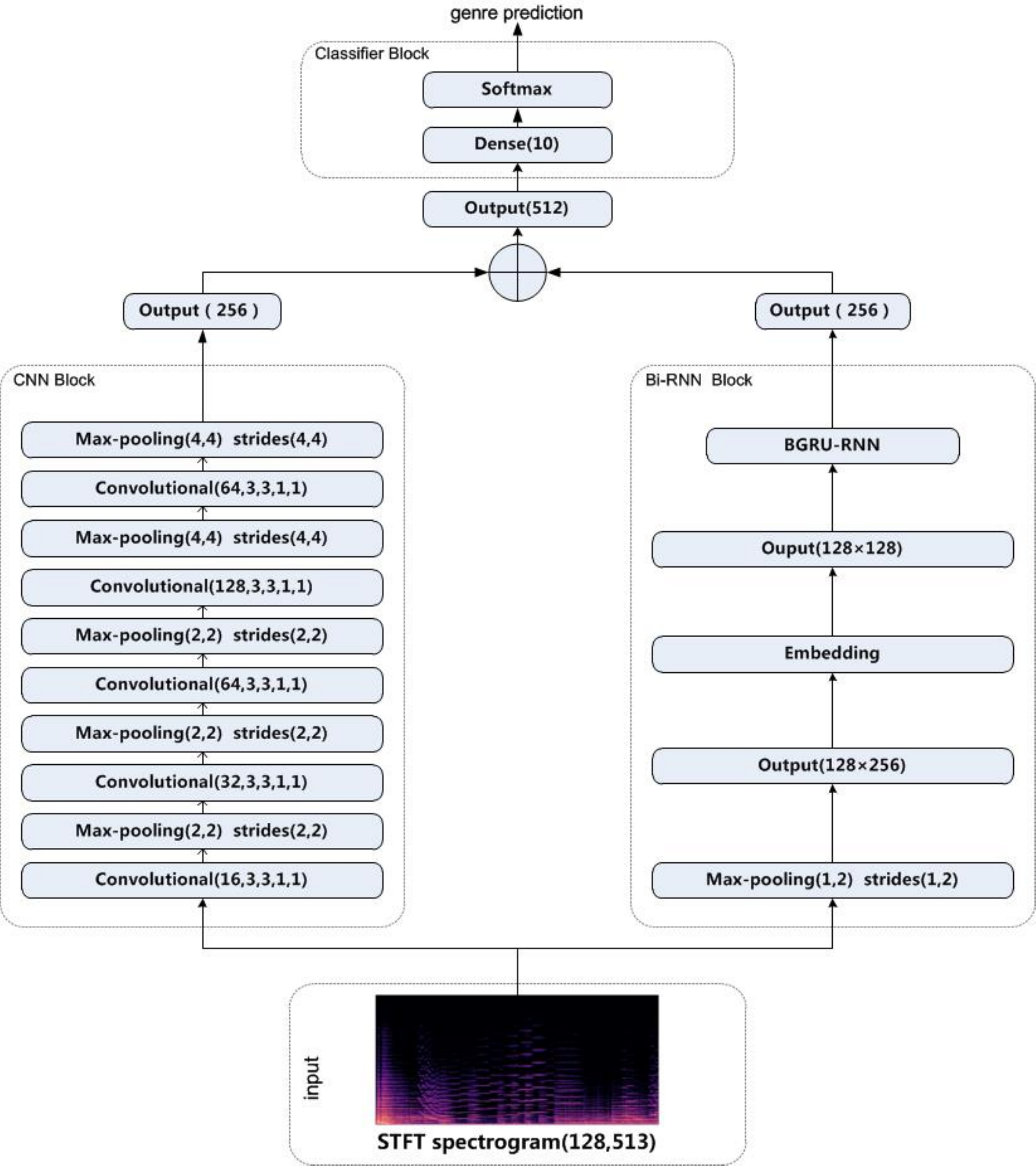}
    \caption{The network architecture of PRCNN}
    \label{fig:my_label}
    \end{center}
    
\end{figure}
As illustrated in Figure 1, our proposed hybrid architecture is divided into four blocks with weights to play different roles. At the bottom of Figure 1, we utilize Short-term Fourier Transform (STFT) spectrogram of musical signals as the input of our network. The input whose size is $128\times513$ is simultaneously fed into paralleling CNN and Bi-RNN blocks to implement feature extraction. As aforementioned, CNNs have excellent performance on extracting spatial features of music. However, the STFT spectrogram of musical signals has some significant sequential-relationships lost in CNNs during supervised learning. Thus, the paralleling Bi-RNN block is employed to extract temporal frame orders from the spectrogram as a supplement. Then the outputs of two paralleling blocks are fused into one feature vector which will be classified next. After a dense layer, we apply a softmax operation as a post-processing stage to acquire a feature vector which consists of normalized probabilities of different music genres. 
As mentioned in Section 1, feature extraction is a crucial part in music genre classifications. Therefore, in the rest of this section, we describe the paralleling CNN and Bi-RNN blocks utilized for feature extraction in detail. 

\subsection{Convolutional Neural Network Block}
Except for the input and output layers, the CNN block of our proposed hybrid architecture has 10 layers, including five convolutional-pooling layers. After each convolutional layer, a max-pooling operation is followed to further process the output of previous convolutional layer. Each kernel detects a fixed $3\times1$ region in the previous layer with $1\times1$ padding. The design of padding is to reduce the information loss during convolution. In order to acquire more meaningful representations from spectrogram, we design the five convolutional layers with 16, 32, 64, 128 and 64 filters respectively. The first three max-pooling layers output the maximum value within a $2\times2$ rectangular neighborhood with strides $2\times2$. And the upper two max-pooling layers reports the maximum value of a $4\times4$ region with $4\times4$ strides to extract more robust representations. The output of CNN block is a vector of $1\times256$ and will be fed into the classifier in conjunction with the output of Bi-RNN block. 

\paragraph{Convolution – kernel size}  Kernels are regarded as feature detectors in convolutional layers. In general, a kernel size defined as $k\times r\times c$ means the kernel can learn k features of $r\times c$, where r and c refers to rows and columns of a kernel respectively. Kernel size determines the range of a feature map it can precisely detects. Thus, the kernel size can certainly affect the performance of feature learning. When the kernel size is too small, it is not capable to learn representative features from the given data. Thus some researchers, such as Krizhevsky et al. \cite{krizhevsky2012imagenet}, proposed large convolution kernels sized as $11\times11\times3$ to detect features. However, the increasing size of convolution kernel makes parameters of per feature detector increase, and obviously, the storage and computation will both increase. Moreover, large kernels lose the invariance within their ranges \cite{choi2016automatic}. Aiming to learn more representative features with less parameters, the kernel size utilized in our proposed architecture is $3\times1$, which have shown excellent performance of features detecting with suitable parameters storage and computation.

\paragraph{Pooling} Pooling function, is regarded as a process of subsampling and a crucial stage in CNNs. In contrast with convolution, pooling is a non-linear behavior which produces a summary statistic of the nearby output. The max-pooling operation employed in CNN block can represent the most prominent features of music, such as amplitudes. A max-pooling can also reduce the dimension of previous output, and therefore prevents the network from overfitting with less parameters. Meanwhile, the pooling size is also an important aspect which influences the music genre classification. In general, undersized pooling size makes the network not invariant enough for some small translations. On the contrary, if the pooling size is oversized, some requisite feature locations will be lost and some error may be brought into the classification result.

\paragraph{Rectified Linear Units} As we all know, convolution is a linear operation which is usually not enough to reflect the representations of features. Thus, we employ Rectified Linear Units (ReLUs) \cite{nair2010rectified} to achieve a non-linear behavior. The definition of ReLUs activation function is $f(x)= max (0,x)$. Obviously, ReLUs brings out sparse feature representations in hidden layers since components below 0 are cut off. In contrast with sigmoid, ReLUs do not saturate at 1 and the partial derivative of the activation function is never 0, which can avoid the appearance of vanishing gradient in some degree. Meanwhile, ReLUs also have more rapid speed of convergence than traditional sigmoid and tanh activations. 

\subsection{Bidirectional Gated Recurrent Units Block}
As illustrated in Figure 1, the BGRU-RNN block consists of 7 layers except for the input and fused output layers. In this block, the input is first processed by a max-pooling layer to reduce the dimension. After this step, the dimension of spectrogram is reduced to $128\times256$. Since the upper BGRU layers are constructed kinder complex, we employ an embedding layer for further dimension reduction to decrease parameters of. After the pre-training, a $128\times128$ input is fed into two stacked BGRUs illustrated in Figure 2 for features extraction. In contrast to the output of CNNs block, we simply splice the outputs of two stacked BGRU layers as one 256D feature vector. 

As we all know, standard recurrent neural networks (RNNs) only take advantage of previous contexts but ignore the backwards dependencies which are also important for feature learning. However, many applications have demonstrated that the prediction of $y^{(t)}$ heavily depends on the whole input sequence, including the past and future information. Another limitation of traditional RNNs is that they will suffer from the problem of vanishing and exploding gradients when dealing(deal) with long-term dependencies. Thus, in our hybrid architecture, we exploit two stacked bidirectional BGRUs which is a variant of RNNs to improve the performance of feature extraction. The structure of BGRUs is shown in Figure 2 and we will describe it in detail soon.
\begin{figure}
    \begin{center}
        \includegraphics[width=10cm]{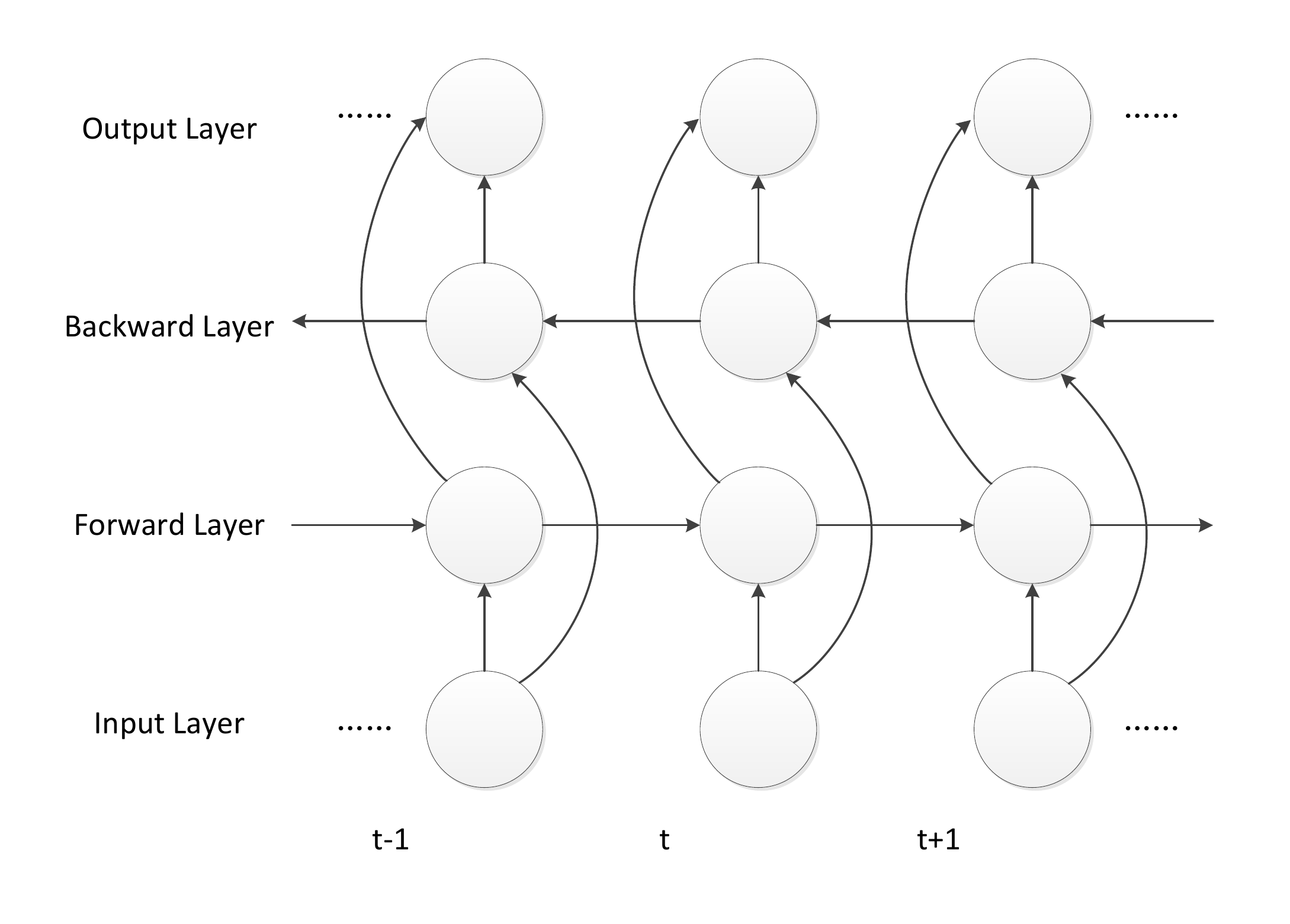}
    \caption{The network architecture of BGRU}
    \label{fig:my_label}
    \end{center}
    
\end{figure}

\paragraph{Bidirectional Gated Recurrent Units}

The design of BGRU is motivated by two main considerations: 1) utilizing gated Recurrent Unit (GRU) to extract temporal features from spectrogram of musical signals which are lost in CNNs; 2) extracting powerful representations by taking full advantage of past and future information of a sequence. 

GRU is proposed in \cite{cho2014properties} to make the recurrent blocks adaptively capture information from variable-length sequences. Obviously, a BGRUs architectures means that we employ GRU in both forward states part and backward states part. As illustrated in Figure 2, the input layer is fed into both forward and backward layers. Meanwhile, the output layer is produced by both forward and backward layers. But the two reverse layers have no direct connections.

Indeed, GRU is a more simplified variation of the Long Short-term Memory (LSTM) \cite{hochreiter1997long}, which integrates input and forget gates into one “update gate” and append a “reset gates”.

For GRU, it makes one single gating unit simultaneously controls the forgetting element and the decision to update the state unit. In the $i-th$ GRU, the activation $h_i^{(t)}$ at time $t$ is calculated by the previous activation $h_i^{(t-1)}$ and the current candidate update:
\begin{equation}
    h_i^{(t)}=u_i^{(t)}\tilde{h}_i^{(t)}+(1-u_i^{(t)})h_i^{(t-1)},
\end{equation}
where $u$ and $\tilde{h}_i^{(t)}$ respectively stand for “update gate” and candidate activation. The update gate decides how much the unit updates from its activation:
    \begin{equation}
        u_i^{(t)}=\sigma(b_u+U_ux^{(t)}+W_uh^{(t-1)})^i,
    \end{equation}
where $b$, $U$ and $W$ respectively denote the biases, input weights and recurrent weights into the $i-th$ GRU. The input vector at time $t$ is defined as $x^{(t)}$. The candidate activation $\tilde{h}_i^{(t)}$ is computed analogously to the update gate:
    \begin{equation}
        \tilde{h}_i^{(t)}=tanh(b+Ux^{(t)}+W(r^{(t)}\otimes h^{(t-1)}))^i,
    \end{equation}
where $r$ stands for “reset gate” and $\otimes$ denotes an element-wise multiplication operation. If $r^{(t)}$ is close to 0, the reset gate is off and the unit should forget the past information. The reset gate is defined with the following formula: 
    \begin{equation}
        r_i^{(t)}=\sigma(b_r+U_rx^{(t)}+W_rh^{(t-1)})^i
    \end{equation}
The update and reset gates can separately “neglect” vector parts. The update gates decide how much the past states should impact current states. While the reset gates provide nonlinear effect in the correlation between past state and future state. They decide which parts should be computed in the future state. 

In our bidirectional architecture, the forward GRUs are calculated by past states along positive time axis while the back forward GRUs are computed by future states along reverse time axis. For instance, the activation at time $t$ of backward GRUs is calculated by the future activation $h_i^{(t+1)}$ and the current candidate update:
    \begin{equation}
        h_i^{(t)}=u_i^{(t)}\tilde{h}_i^{(t)}+(1-u_i^{(t)})h_i^{(t+1)},
    \end{equation}
and other formulas are similar to this, being computed along the reverse time axis.

Comparing with LSTM, GRU has simpler structure which captures temporal correlations from musical signals but overcomes the problem of vanishing and exploding gradient. GRU and LSTM can both preserve important information via gates inside during dealing with long-term dependencies. But in GRU, the activations of gates only depend on previous output and current input. Thus, the simpler GRU mitigates the occurrence of overfitting and tends to converge faster than LSTM with less parameters.

\subsection{Feature Fusion and Classifier Block}
The outputs of the two paralleling blocks are two 256 dimensional vectors. In our hybrid architecture, CNNs and BRNNs blocks respectively focus on extracting spatial features and temporal frame orders of musical signals. Thus, the two vectors need to be fused into one powerful representation to improve the performance of music genre classification. Since the two vectors have the same size, we carry out two methods of fusing them into one feature representation: 1) directly add the values of two vectors together and acquire a new 521 dimensional vector; 2) keep the original values of two vectors and concatenate them into a 521 dimensional vector. After feature fusion, the syncretic representation is fed into dense and softmax layers to implement the classification.

In the classifier block, a dense layer is employed to map the previous fused vector into a feature vector whose size is 10. Then a softmax function is adopted in this feature vector for music genre classification. The softmax function is defined as:
    \begin{equation}
        P(i)=\frac{exp(x_i)}{\sum_{k=1}^kexp(x_k)},
    \end{equation}
where $P(i)$ and $x_i$ respectively represent the probability of music genre and the $i-th$ value of the feature vector. The aim of exploiting a softmax function is to make each value of feature vector between 0~1. And the result of $\sum_{k=1}^kexp(x_k)$ equals 1. In this situation, the 10 values between 0~1 can be regarded as the probabilities of 10 music genres.

\section{EXPERIMENT}
In this section, we introduce the two dataset used in our experiments and report some contrast experiments results for validating the effectiveness of the proposed paralleling architecture. 
\subsection{Dataset Description}
There are two classical datasets utilized in our experiments. One is GTZAN dataset \cite{tzanetakis2002musical} which has been used as a benchmark in various systems for music genre classification. It consists of 1000 songs excerpts which are evenly distributed into ten different genres: Blues, Classical, Country, Disco, Hippop, Jazz, Metal, Pop, Reggae and Rock. Each song is about 30 seconds duration and sampled with the rate of 22050Hz at 16 bit. 

Another dataset is Extended Ballroom dataset \cite{marchand2016extended} which is an extended version based on Ballroom dataset \cite{Gouyon2004Evaluating}. The Extended Ballroom dataset we use for training and testing consists of 4180 excerpts with 30 seconds duration. The audio quality is better than the Ballroom dataset and 5 new genres of ballroom dance music: Foxtrot, Pasodoble, Salsa, Slowwaltz and Wcswing are  added.

\subsection{Experimental Setup}
\paragraph{Dataset pre-processing}

As we all know, Deep Neural Networks need enormous input data to learn robust feature representation. However, the datasets we used in our experiments are with 1000 song excerpts and 4180 music tracks respectively. In order to increase the number of tracks, we cut each song excerpt into shorter music clips with 3 seconds duration and 50\% overlap. Thus, the increased training datasets help our architecture avoid overfitting partly and have better performance on feature extraction. Similar to the processing in \cite{sigtia2014improved}\cite{zhang2016improved}, we calculate Fast Fourier Transforms (FFTs) on frames of length 1024 at 22050 kHz sampling rate with 50\% overlap and use the absolute value of each FFT frame. We finally construct a STFT spectrogram with 128 frames and each frame is a 513 dimensional vector.

\subsection{Result}
\begin{table}[]
    \centering
    \setlength{\belowcaptionskip}{10pt}%
    \caption{Genre classification results on GTZAN dataset}
    \begin{tabular}{ccc}
    \hline
    \textbf{Methods} & \textbf{Features} & \textbf{Accuracy} \\
        \hline
        \textbf{CNN+2-layer RNN} & \textbf{STFT} & \textbf{88.8\%} \\
        \textbf{CNN+1-layer RNN} & \textbf{STFT} & \textbf{90.2\%} \\
        nnet1 & STFT & 84.8\% \\
        nnet2 & STFT 87.4\% \\
        KCNN(k=5)+SVM \cite{zhang2015deep} & Mel-spectrum, SFM, SCF & 83.9\% \\
        DNN(ReLU+SGD+Dropout) \cite{sigtia2014improved} & FFT(aggregation) & 83.0\% \\
        Multilayer invariant representation \cite{zhang2014deep} & STFT with log representation & 82.0\% \\
        \hline
    \end{tabular}
    \label{tab:my_label}
\end{table}

The music genre classification accuracy of the proposed PRCNN is reported in Table 1. For comparison, we also reported other achievements applied to the GTZAN dataset presented in \cite{zhang2016improved}. As shown in Table 1, we design our Bi-RNN block with 2 layers RNNs and 1 layer RNN respectively. And the results both show better performance than other achievements applied to the same dataset. Nevertheless, the problem of overfitting can easily appears in 2 layers RNNs during feature learning in the small sized dataset. Thus, we only use 1 layer RNN in our Bi-RNN block to extract features from spectrogram. And the results prove that the Bi-RNN block with 1 layer RNN achieves better performance than employing 2 layers RNNs.

\begin{table}[]
    \centering
    \setlength{\belowcaptionskip}{10pt}%
    \caption{Improved performance with RNN for different CNNs}
    \begin{tabular}{ccc}
    \hline
        \textbf{CNNs} & \textbf{Without RNN} & \textbf{With RNN} \\
        \hline
        Our CNN & 88.0\% & \textbf{92.0\%} \\
        Alexnet & 81.4\% & \textbf{88.8\%} \\
        Vgg11 & 86.8\% & \textbf{88.7\%} \\
        ResNet-11 & 86.8\% & \textbf{87.6\%} \\
        \hline
    \end{tabular}
    
    \label{tab:my_label}
\end{table}

In order to validate the effectiveness of the additionally paralleling RNN block, we design some contrast experiments with other typical CNNs. In Table 2, all the results are all achieved on the GTZAN dataset. And as can be seen, in contrast to utilizing CNNs alone, all of the CNNs with paralleling RNN can improve the performance of music genre classification.  

\section{Conclusion}
In this paper, we propose a hybrid architecture PRCNN to improve the performance of music genre classification. This end-to-end model consists of paralleling CNN and Bi-RNN blocks for feature extraction. The CNN block focuses on extracting spatial features from spectrogram of musical signals. On the contrary, the BRNNs block is designed with the purpose of modeling temporal frame orders. Furthermore, the bidirectional architecture can make current states depend on not only previous information but also future contexts of the sequence during supervised learning. The outputs of two paralleling blocks are fused into a more powerful feature vector for music classification. Several experiments in this paper adequately demonstrate the effectiveness of our hybrid architecture. Moreover, comparing with utilizing CNNs alone, the experimental results prove extracting temporal frame orders from musical signals with RNNs improves the performance of music genre classification.


\begin{thebibliography}{10}

\bibitem{tzanetakis2002musical}
G.~Tzanetakis and P.~Cook, ``Musical genre classification of audio signals,''
  {\em IEEE Transactions on speech and audio processing}, vol.~10, no.~5,
  pp.~293--302, 2002.

\bibitem{shawe2005investigation}
J.~Shawe-Taylor and A.~Meng, ``An investigation of feature models for music
  genre classification using the support vector classifier,'' 2005.

\bibitem{west2005finding}
K.~West and S.~Cox, ``Finding an optimal segmentation for audio genre
  classification.,'' in {\em ISMIR}, pp.~680--685, 2005.

\bibitem{Duda2000Pattern}
R.~O. Duda, P.~E. Hart, and D.~G. Stork, ``Pattern classification (2nd
  edition),'' {\em En Broeck the Statistical Mechanics of Learning Rsity},
  2000.

\bibitem{fu2011survey}
Z.~Fu, G.~Lu, K.~M. Ting, and D.~Zhang, ``A survey of audio-based music
  classification and annotation,'' {\em IEEE transactions on multimedia},
  vol.~13, no.~2, pp.~303--319, 2011.

\bibitem{baniya2014novel}
B.~K. Baniya, D.~Ghimire, and J.~Lee, ``A novel approach of automatic music
  genre classification based on timbrai texture and rhythmic content
  features,'' in {\em Advanced Communication Technology (ICACT), 2014 16th
  International Conference on}, pp.~96--102, IEEE, 2014.

\bibitem{huang2006extreme}
G.-B. Huang, Q.-Y. Zhu, and C.-K. Siew, ``Extreme learning machine: theory and
  applications,'' {\em Neurocomputing}, vol.~70, no.~1, pp.~489--501, 2006.

\bibitem{breiman1996bagging}
L.~Breiman, ``Bagging predictors,'' {\em Machine learning}, vol.~24, no.~2,
  pp.~123--140, 1996.

\bibitem{arabi2009enhanced}
A.~F. Arabi and G.~Lu, ``Enhanced polyphonic music genre classification using
  high level features,'' in {\em Signal and Image Processing Applications
  (ICSIPA), 2009 IEEE International Conference on}, pp.~101--106, IEEE, 2009.

\bibitem{sarkar2015music}
R.~Sarkar and S.~K. Saha, ``Music genre classification using emd and pitch
  based feature,'' in {\em Advances in Pattern Recognition (ICAPR), 2015 Eighth
  International Conference on}, pp.~1--6, IEEE, 2015.

\bibitem{wei2016hcp}
Y.~Wei, W.~Xia, M.~Lin, J.~Huang, B.~Ni, J.~Dong, Y.~Zhao, and S.~Yan, ``Hcp: A
  flexible cnn framework for multi-label image classification,'' {\em IEEE
  transactions on pattern analysis and machine intelligence}, vol.~38, no.~9,
  pp.~1901--1907, 2016.

\bibitem{ciresan2011flexible}
D.~C. Ciresan, U.~Meier, J.~Masci, L.~Maria~Gambardella, and J.~Schmidhuber,
  ``Flexible, high performance convolutional neural networks for image
  classification,'' in {\em IJCAI Proceedings-International Joint Conference on
  Artificial Intelligence}, vol.~22, p.~1237, Barcelona, Spain, 2011.

\bibitem{dieleman2014end}
S.~Dieleman and B.~Schrauwen, ``End-to-end learning for music audio,'' in {\em
  Acoustics, Speech and Signal Processing (ICASSP), 2014 IEEE International
  Conference on}, pp.~6964--6968, IEEE, 2014.

\bibitem{li2010automatic}
T.~L. Li, A.~B. Chan, and A.~Chun, ``Automatic musical pattern feature
  extraction using convolutional neural network,'' in {\em Proc. Int. Conf.
  Data Mining and Applications}, 2010.

\bibitem{zhang2016improved}
W.~Zhang, W.~Lei, X.~Xu, and X.~Xing, ``Improved music genre classification
  with convolutional neural networks.,'' in {\em INTERSPEECH}, pp.~3304--3308,
  2016.

\bibitem{elman1990finding}
J.~L. Elman, ``Finding structure in time,'' {\em Cognitive science}, vol.~14,
  no.~2, pp.~179--211, 1990.

\bibitem{pons2016experimenting}
J.~Pons, T.~Lidy, and X.~Serra, ``Experimenting with musically motivated
  convolutional neural networks,'' in {\em Content-Based Multimedia Indexing
  (CBMI), 2016 14th International Workshop on}, pp.~1--6, IEEE, 2016.

\bibitem{schuster1997bidirectional}
M.~Schuster and K.~K. Paliwal, ``Bidirectional recurrent neural networks,''
  {\em IEEE Transactions on Signal Processing}, vol.~45, no.~11,
  pp.~2673--2681, 1997.

\bibitem{lawrence1997face}
S.~Lawrence, C.~L. Giles, A.~C. Tsoi, and A.~D. Back, ``Face recognition: A
  convolutional neural-network approach,'' {\em IEEE transactions on neural
  networks}, vol.~8, no.~1, pp.~98--113, 1997.

\bibitem{he2016deep}
K.~He, X.~Zhang, S.~Ren, and J.~Sun, ``Deep residual learning for image
  recognition,'' in {\em Proceedings of the IEEE conference on computer vision
  and pattern recognition}, pp.~770--778, 2016.

\bibitem{choi2016convolutional}
K.~Choi, G.~Fazekas, M.~Sandler, and K.~Cho, ``Convolutional recurrent neural
  networks for music classification,'' {\em arXiv preprint arXiv:1609.04243},
  2016.

\bibitem{krizhevsky2012imagenet}
A.~Krizhevsky, I.~Sutskever, and G.~E. Hinton, ``Imagenet classification with
  deep convolutional neural networks,'' in {\em Advances in neural information
  processing systems}, pp.~1097--1105, 2012.

\bibitem{choi2016automatic}
K.~Choi, G.~Fazekas, and M.~Sandler, ``Automatic tagging using deep
  convolutional neural networks,'' {\em arXiv preprint arXiv:1606.00298}, 2016.

\bibitem{nair2010rectified}
V.~Nair and G.~E. Hinton, ``Rectified linear units improve restricted boltzmann
  machines,'' in {\em Proceedings of the 27th international conference on
  machine learning (ICML-10)}, pp.~807--814, 2010.

\bibitem{cho2014properties}
K.~Cho, B.~Van~Merri{\"e}nboer, D.~Bahdanau, and Y.~Bengio, ``On the properties
  of neural machine translation: Encoder-decoder approaches,'' {\em arXiv
  preprint arXiv:1409.1259}, 2014.

\bibitem{hochreiter1997long}
S.~Hochreiter and J.~Schmidhuber, ``Long short-term memory,'' {\em Neural
  computation}, vol.~9, no.~8, pp.~1735--1780, 1997.

\bibitem{marchand2016extended}
U.~Marchand and G.~Peeters, ``The extended ballroom dataset,'' 2016.

\bibitem{Gouyon2004Evaluating}
F.~Gouyon, S.~Dixon, E.~Pampalk, and G.~Widmer, ``Evaluating rhythmic
  descriptors for musical genre classification,'' 2004.

\bibitem{sigtia2014improved}
S.~Sigtia and S.~Dixon, ``Improved music feature learning with deep neural
  networks,'' in {\em Acoustics, Speech and Signal Processing (ICASSP), 2014
  IEEE International Conference on}, pp.~6959--6963, IEEE, 2014.

\bibitem{zhang2015deep}
P.~Zhang, X.~Zheng, W.~Zhang, S.~Li, S.~Qian, W.~He, S.~Zhang, and Z.~Wang, ``A
  deep neural network for modeling music,'' in {\em Proceedings of the 5th ACM
  on International Conference on Multimedia Retrieval}, pp.~379--386, ACM,
  2015.

\bibitem{zhang2014deep}
C.~Zhang, G.~Evangelopoulos, S.~Voinea, L.~Rosasco, and T.~Poggio, ``A deep
  representation for invariance and music classification,'' in {\em Acoustics,
  Speech and Signal Processing (ICASSP), 2014 IEEE International Conference
  on}, pp.~6984--6988, IEEE, 2014.

\end{thebibliography}
\end{document}